\documentclass{jfm}
\usepackage{graphicx}
\usepackage{epstopdf, epsfig}
\usepackage{upgreek} 
\usepackage{bm} 
\usepackage{mathtools} 
\usepackage{mathrsfs}
\usepackage{lineno}
\usepackage{color}
\usepackage{xcolor}
\usepackage{tabularx}
\usepackage[abs]{overpic}
\usepackage{placeins}
\usepackage{tikz}
\usepackage{subfig}

\usepackage[colorlinks,
            linkcolor=olive,
            anchorcolor=green,
            citecolor=teal
            ]{hyperref}

\shorttitle{Mechanisms governing the motion of settling particles in wall-bounded turbulence}
\shortauthor{A.D. Bragg, D.H. Richter, and G. Wang}

\title{Mechanisms governing the settling velocities and spatial distributions of inertial particles in wall-bounded turbulence}

\author{
    A.D. Bragg\corresp{\email{andrew.bragg@duke.edu}}\aff{1},
    D. H. Richter\aff{2}
  \and G. Wang\aff{3}}
  
    \affiliation{
\aff{1} Department of Civil and Environmental Engineering, Duke University, Durham, NC 27708, USA
\aff{2}Department of Civil and Environmental Engineering and Earth Sciences, University of Notre Dame, Notre Dame, Indiana 46556, USA
\aff{3} Physics of Fluids Group and Twente Max Planck Center, Department of Science and Technology, Mesa+ Institute, and J. M. Burgers Center for Fluid Dynamics, University of Twente, P.O. Box 217, 7500 AE Enschede, The Netherlands
}

\begin{document}

\maketitle

\begin{abstract}

We use theory and Direct Numerical Simulations (DNS) to explore the average vertical velocities and spatial distributions of inertial particles settling in a wall-bounded turbulent flow. The theory is based on the exact phase-space equation for the Probability Density Function describing particle positions and velocities. This allowed us to identify the distinct physical mechanisms governing the particle transport. We then examined the asymptotic behavior of the particle motion in the near-wall region, revealing the fundamental differences to the near wall behavior that is produced when incorporating gravitational settling. When the average vertical particle mass flux is zero, the averaged vertical particle velocity is zero away from the wall due to the particles preferentially sampling regions where the fluid velocity is positive, which balances with the downward Stokes settling velocity. When the average mass flux is negative, the combined effects of turbulence and particle inertia lead to average vertical particle velocities that can significantly exceed the Stokes settling velocity, by as much as ten times. Sufficiently far from the wall, the enhanced vertical velocities are due to the preferential sweeping mechanism. However, as the particles approach the wall, the contribution from the preferential sweeping mechanism becomes small, and a downward contribution from the turbophoretic velocity dominates the behavior. Close to the wall, the particle concentration grows as a power-law, but the nature of this power law depends on the particle Stokes number. Finally, our results highlight how the Rouse model of particle concentration is to be modified for particles with finite inertia.

\end{abstract}

\section{Introduction}\label{sec:Introduction}

It is well-known that in homogeneous, isotropic turbulence (HIT), small particles with non-negligible inertia will settle at a rate that can exceed their Stokes settling velocity. \citet{maxey1987gravitational} and \citet{wang1993settling} were among the first to outline and characterize the so-called preferential sweeping mechanism, according to which particles falling under the influence of gravity are swept around the downward side of eddies in HIT. This biased sampling of the background velocity field often leads to an average downwards particle velocity that is larger than that if the particles were settling in a quiescent medium --- an effect which has been seen in direct numerical simulation (DNS) \citep{wang1993settling,YangJFM1998} and experiments \citep{aliseda02,YangPoF2003,YangJFM2005,petersen19}. It has even been observed in the settling of snowflakes through the atmospheric boundary layer \citep{nemes2017snowflakes}. 

Accordingly, much effort has been directed at understanding the nature of this effect. It is generally accepted that the maximum observed modification to the settling velocity occurs at a Stokes number (based on the Kolmogorov scales) of order unity, and parameterizations have been developed to describe this \citep{rosa2016settling}. However, there are still questions regarding the magnitude of this effect, as well as how this magnitude depends on other nondimensional parameters such as the Froude number or Reynolds number (see for example the discussion in \citet{nemes2017snowflakes}). The work of \citet{good14} nicely maps out the various settling regimes, and also addresses the possibility of turbulence retarding the settling velocity of inertial particles and how this relates to the drag law in DNS. The recent study of \citet{tom19} advanced the work of \citet{maxey1987gravitational} by developing a theoretical framework that is valid for arbitrary particle inertia and reveals the contribution that different turbulent scales make to the enhanced settling, and how this depends on Stokes number, Froude number, and flow Reynolds number. The theory predicts that for particles with finite inertia, the Reynolds number dependence will always saturate, and that the saturation Reynolds number is a non-decreasing function of particle inertia. These predictions were confirmed by DNS results.

While all of the aforementioned studies have focused on HIT, the picture is significantly less clear in the context of wall-bounded turbulent flows. Much work has been aimed at understanding the interaction of coherent structures with inertial particles, including transport to/from the wall \citep{marchioli02,richter2014modification}, and the turbophoretic drift \citep{reeks83,sardina2012wall,johnson20}; however, the vast majority of these studies neglect wall-normal gravity. Indeed, discrepancies in certain particle velocity statistics between experimental setups with horizontal \citep{RighettiJFM2004,KigerJoT2002,LiEF2012} versus vertical \citep{kulick1994particle,fong2019velocity} channels are possibly due to differences in gravitational orientation.  The recent DNS of \citet{lee2019effect} demonstrates that for two-way coupled flows, the addition of wall-normal gravity can even qualitatively alter the interaction between inertial particles and near-wall streaks and coherent structures via mechanisms similar to that of \citet{wang1993settling}.

Generally speaking, the question of inertial particle settling through wall-bounded turbulence has been largely avoided, and is complicated by several factors as compared to HIT. First, the process of turbophoresis, which is absent in HIT due to a lack of mean gradient in turbulence kinetic energy, is difficult to distinguish from the preferential sweeping mechanism, although they are distinct phenomena. Secondly, the smallest time scales of the flow actually vary with wall-normal location, so it should not be expected that the settling rate is simply a function only of a single Stokes number. The velocity of a particle falling through a flow with spatially varying time scales be a function of wall-normal distance, and will be influenced by the flow Reynolds number as well. Third, rigorous phase-space probability density function theories of particle transport in inhomogeneous turbulence suggest that there is an additional drift effect that contributes to the wall-normal particle motion, which arises from the inhomogeneity but is distinct from the turbophoretic drift, and exists even in the absence of gravitational settling \citep{reeks91,swailes97,bragg12b,bragg14b}. The importance of this additional drift compared with the gravitational and turbophoretic drifts is not well understood, and it is difficult to develop closed expressions that capture its influence \citep{bragg12}. It is one of the goals of this work to distinguish between these transport mechanisms and determine their relative importance.

The settling velocity of inertial particles in wall-bounded flows is also important because it controls the spatial distribution of the particles. In an effort to theoretically extend the logarithmic profile of wall turbulence for passive scalars, \citet{rouse37} derived the well-known power law for the average concentration when the scalar experiences gravitational settling towards the wall, where the power is proportional to the Stokes settling velocity of the particle. This theory, which is valid only within the logarithmic region of the turbulent boundary layer, assumes that gravitational settling is balanced by turbulent fluxes on average, so the net flux is zero at any height and the magnitude of the downward settling flux is equal to the upward turbulent flux. This so-called flux-profile relationship between the mean concentration and flux is the basis for many geophysical measurements which attempt to estimate the surface emission of discrete particles like snow, dust, or water droplets \citep{SommerfeldJGR1965,HoppelJGR2002,LewisSchwartz}, as well as for determining how to specify boundary conditions for heavy particles in coarse-scale numerical models \citep{ChameckiAS2009}. 

Since then, further modifications to the theory of \citet{rouse37} have been made, such as incorporating a net imbalance between the gravitational and turbulent fluxes \citep{kind1992one,HoppelJGR2002}. \citet{freire2016flux} extended the theory to non-neutral stability, including the effects of both stable and unstable stratification, while \citet{nissanka2018parameterized} considered the full boundary layer (i.e. not restricted only to the surface layer). Other studies, such as those by \citet{PanBLM2013}, and \citet{ZhuEFM2017}, explore beyond the one-dimensional framework by incorporating a streamwise dependence in addition to height. Recently, \citet{RichterBLM2018} attempted to incorporate inertial effects into the theoretical framework of \citet{rouse37} by using a first-order perturbation expansion of the particle advection velocity with Stokes number in the equation for mean particle concentration \citep{maxey1987gravitational,druzhinin1995two,ferry2001fast}. This inertial correction to the particle advection velocity provided concentration profiles that matched DNS results, but only in the limit of small Stokes numbers, as expected. For finite particle inertia, the underlying framework of \citet{rouse37} is called into question, because the equation ignores the contributions from a number of different mechanisms that are crucial when the Stokes number is not small. 

In this study, therefore, we set out to understand and quantify the relevant transport mechanisms for inertial particles settling through a wall-bounded turbulent flow. This is done by examining the problem through a phase-space probability density formulation, where the magnitudes and regimes of relevance for the various distinct inertial effects are calculated via DNS. The goal is to clarify the multiple pathways through which inertia can affect the flux-profile relationship in wall-bounded turbulence, in order to clear the way towards more theoretically sound extensions to the theory of \citet{rouse37} in the future. Section \ref{sec:Theory} outlines the theoretical framework by casting the problem in phase space, and section \ref{sec:DNS} describes the DNS used to generate the data. We then discuss the results in section \ref{sec:Results}.

\section{Theory}\label{sec:Theory}

We consider particles with particle-to-fluid density ratio $\rho_p/\rho_f \gg 1$, and whose size is smaller than the smallest length scales of the wall-bounded turbulence. Furthermore, we also consider mass and volume loadings such that the one-way coupled, dilute regime applies. In this case, the equations of motion are
\begin{align}
d\bm{x}^p(t)&=dt\bm{v}^p(t)+\sqrt{2 \kappa dt} d\bm{\xi}(t),\label{eomx}\\
\frac{d}{dt}\bm{v}^p(t)&=\frac{\Psi}{\tau_p}\Big(\bm{U}^p(t)-\bm{v}^p(t)\Big)+\bm{g},\label{eomv}
\end{align}
where $\bm{x}^p(t),\bm{v}^p(t)$ are the particle position and velocity, $\tau_p$ is the particle response time, $\bm{U}^p(t)\equiv\bm{U}(\bm{x}^p(t),t)$ corresponds to the fluid velocity field evaluated at the particle position, $\kappa$ is a constant diffusion coefficient, and $\bm{g}$ is gravity. The term $d\bm{\xi}(t)$ is a normalized vector-valued Wiener process with unit variance. This diffusion term is included since it will be used in the DNS in order to enable the particles to be suspended from the wall and generate a configuration where the vertical particle mass flux is zero (to be described in more detail below).

The variable $\Psi\equiv [1+0.15(Re _p)^{0.687}]$ appears due to using the Schiller-Naumann \citep{schiller1933ber} hydrodynamic drag force model, where $Re_p$ is the particle Reynolds number. In the theoretical developments below, for analytical tractability we assume $Re_p\to 0$ and hence take $\Psi=1$. In the DNS this assumption will not be made, however, the DNS results show that on average $Re_p<1$. As a result, the assumption that $\Psi=1$ in the theory will only lead to minor differences between the theory and DNS. 

We consider the particle motion in a phase-space $\bm{x},\bm{v}$ with Probability Density Function (PDF)
\begin{align}
\mathcal{P}(\bm{x},\bm{v},t)\equiv\Big\langle\delta(\bm{x}^p(t)-\bm{x})\delta(\bm{v}^p(t)-\bm{v})\Big\rangle,
\end{align}
that satisfies $\int_{\mathbb{R}^3}\int_{\Omega}\mathcal{P}(\bm{x},\bm{v},t)\,d\bm{x}\,d\bm{v}=1$, where $\Omega\subset\mathbb{R}^3$ denotes the domain of the flow. Here, $\langle\cdot\rangle$ denotes an ensemble average over all realizations of the system, and $\delta()$ denotes a Dirac distribution. The exact PDE governing $\mathcal{P}$ is \citep{bragg14d}
\begin{align}
\begin{split}
\partial_t\mathcal{P}+\bm{v\cdot\nabla_x}\mathcal{P}=& \kappa\bm{\nabla_x}^2\mathcal{P}- \frac{1}{\tau_p} \bm{\nabla_v \cdot}\Big(\langle\bm{U}\rangle\mathcal{P}\Big)\\
&-\frac{1}{\tau_p} \bm{\nabla_v \cdot}\Big(\mathcal{P}\langle \bm{u}^p(t)\rangle_{\bm{x},\bm{v}}\Big)+\frac{1}{\tau_p} \bm{\nabla_v \cdot}\Big(\bm{v}\mathcal{P}\Big)-\bm{g\cdot\nabla_v  }\mathcal{P},
\end{split}
\label{PDFeq}
\end{align}
where $\bm{\nabla_x}^2=\bm{\nabla_x \cdot\nabla_x}$, the operator $\langle\cdot\rangle_{\bm{x},\bm{v}}$ denotes an ensemble average conditioned on $\bm{x}^p(t)=\bm{x}$, $\bm{v}^p(t)=\bm{v}$, and
\begin{align}
\bm{u}^p(t)\equiv\bm{U}^p(t) -\langle\bm{U}(\bm{x},t)\rangle\Big\vert_{\bm{x}=\bm{x}^p(t)}.
\end{align}
From \eqref{PDFeq}, the transport equations governing the moments of the PDF may be constructed. The zeroth moment obeys the equation
\begin{align}
\frac{D}{Dt}\varrho&=-\varrho\bm{\nabla_x\cdot}\langle\bm{v}^p(t)\rangle_{\bm{x}}+\kappa\bm{\nabla_x}^2\varrho,\label{M0}
\end{align}
where
\begin{align}
\frac{D}{Dt}&\equiv \partial_t +\langle\bm{v}^p(t)\rangle_{\bm{x}}\bm{\cdot\nabla_x},
\end{align}
the operator $\langle\cdot\rangle_{\bm{x}}$ denotes an ensemble average conditioned on $\bm{x}^p(t)=\bm{x}$, and
\begin{align}
\varrho(\bm{x},t)\equiv\int_{\mathbb{R}^3}\mathcal{P}(\bm{x},\bm{v},t)\,d\bm{v},
\end{align}
is the marginal PDF that describes the spatial distribution of the particles. 

The first moment describes the momentum of the particle phase and is governed by
\begin{align}
\varrho\frac{D}{Dt}\langle\bm{v}^p(t)\rangle_{\bm{x}}&=\kappa\bm{\nabla_x}^2\varrho\langle\bm{v}^p(t)\rangle_{\bm{x}}-\bm{\nabla_x\cdot}\varrho\bm{S}+\frac{\varrho}{\tau_p}\langle\bm{u}^p(t)\rangle_{\bm{x}}+\frac{\varrho}{\tau_p}\langle\bm{U}\rangle-\frac{\varrho}{\tau_p}\langle\bm{v}^p(t)\rangle_{\bm{x}}+\varrho\bm{g},\\
\bm{S}(\bm{x},t)&\equiv\Big\langle\Big(\bm{v}^p(t)-\langle\bm{v}^p(t)\rangle_{\bm{x}}\Big)\Big(\bm{v}^p(t)-\langle\bm{v}^p(t)\rangle_{\bm{x}}\Big)  \Big\rangle_{\bm{x}},
\end{align}
where $\bm{S}$ is the particle fluctuating velocity covariance tensor. We may also re-write this equation as an expression for the mean particle velocity 
\begin{align}
\langle\bm{v}^p(t)\rangle_{\bm{x}}=-\tau_p\frac{D}{Dt}\langle\bm{v}^p(t)\rangle_{\bm{x}}+\frac{\tau_p \kappa}{\varrho}\bm{\nabla_x}^2\varrho\langle\bm{v}^p(t)\rangle_{\bm{x}}-\frac{\tau_p}{\varrho}\bm{\nabla_x\cdot}\varrho\bm{S}+\langle\bm{u}^p(t)\rangle_{\bm{x}}+\langle\bm{U}\rangle+\tau_p\bm{g}.\label{MPV}
\end{align}
These transport equations for $\varrho$ and $\langle\bm{v}^p(t)\rangle_{\bm{x}}$ are exact, and do not introduce any approximations beyond those already contained in the equations of motion \eqref{eomx} and \eqref{eomv} themselves.

\subsection{Vertical transport in a stationary, wall-bounded turbulent flow}
In this study we consider a horizontal, statistically stationary turbulent channel flow, and denote by $\bm{e}_z$ the unit vector in the vertical direction, with $\bm{g}=-g\bm{e}_z$, $\bm{x\cdot}\bm{e}_z=z$, $\bm{e}_z\bm{\cdot\nabla_x}=\nabla_z$,  $\bm{v}^p(t)\bm{\cdot}\bm{e}_z=w^p(t)$, $\bm{u}^p(t)\bm{\cdot}\bm{e}_z=u^p(t)$, $\langle\bm{U}\rangle\bm{\cdot}\bm{e}_z=0$. We then obtain from \eqref{M0} and \eqref{MPV}
\begin{align}
\varrho&=(\Phi+\kappa\nabla_z\varrho)\langle{w}^p(t)\rangle_{z},\label{VF}
\end{align}
\begin{align}
\langle{w}^p(t)\rangle_{{z}}&=\underbrace{-\frac{\tau_p}{2}\nabla_z \langle w^p(t)\rangle^2_{{z}}}_{\text{R1}}-\underbrace{\frac{\tau_p}{\varrho} S\nabla_z \varrho}_{\text{R2}}-\underbrace{\tau_p\nabla_z S}_{\text{R3}}+\underbrace{\langle{u}^p(t)\rangle_{{z}}}_{\text{R4}}-\underbrace{\tau_p g}_{\text{R5}}+\underbrace{\frac{\tau_p \kappa}{\varrho}\nabla_z^2\varrho\langle{w}^p(t)\rangle_{z}}_{\text{R6}},\label{rhoeq}
\end{align}
where $\Phi$ is an integration constant that is determined by the boundary conditions and corresponds to the total net mass flux, and $S=\bm{S:}(\bm{e}_z\bm{e}_z)$. These equations are unclosed, both due to $\langle{u}^p(t)\rangle_{{z}}$ and $S$.

Equation \eqref{VF} is singularly perturbed with respect to $\kappa$, and therefore for $\kappa\neq 0$ the solution to \eqref{VF} is
\begin{align}
\varrho&=\frac{\Phi}{\kappa}\int_0^z \mathcal{I}(z,y)\,dy,\\
\mathcal{I}(z,y)&\equiv \exp\Bigg( \frac{1}{\kappa}\int_y^z \langle{w}^p(t)\rangle_{{q}}\,dq\Bigg),
\end{align}
while for $\kappa=0$ we have simply $\varrho =\Phi/\langle{w}^p(t)\rangle_{z}$. In either case, this highlights that understanding $\langle{w}^p(t)\rangle_{z}$ is important not only to quantify the average settling velocity of the particles, but also because it determines the distribution $\varrho(z)$ for a given mass flux $\Phi$.

We introduce the local Stokes number $St\equiv\tau_p/\tau_L$, where $\tau_L(z)$ is the fluid integral timescale at height $z$ from the wall, and the Froude number $Fr\equiv\langle\|\bm{a}\|^2\rangle^{1/2}/g$, where $\bm{a}$ is the fluid acceleration. In the limit $St\to 0$ with finite $St/Fr$, \eqref{rhoeq} reduces to $\langle{w}^p(t)\rangle_{{z}}=\langle{u}^p(t)\rangle_{{z}}-\tau_p g$. With this asymptotic behavior, then for the zero-flux case $\Phi=0$, we obtain from \eqref{VF}
\begin{align}
0=\varrho\langle{u}^p(t)\rangle_{{z}}-\tau_p\varrho g-\kappa\nabla_z\varrho.\label{rhoeq2}
\end{align}
Using a gradient-diffusion approximation $\varrho\langle{u}^p(t)\rangle_{{z}}\approx -\mathcal{K}\nabla_z\varrho$ leads to
\begin{align}
0=-(\mathcal{K}+\kappa)\nabla_z\varrho-\tau_p\varrho g.\label{Rouse_rho}
\end{align}
The result in \eqref{Rouse_rho} corresponds to the phenomenological model for $\varrho$ by \cite{rouse37} when the diffusion coefficient $\mathcal{K}$ is based on an eddy-diffusivity approximation for the log-law region of a wall-bounded turbulent flow. As the analysis above shows, \eqref{Rouse_rho} is restricted to the limit $St\to 0$ with $St/Fr$ finite, and an important challenge is how the Rouse model is to be extended to capture the effects of finite particle inertia, as attempted in \citet{RichterBLM2018}. Moreover, there is uncertainty regarding the validity of the gradient-diffusion closure $\varrho\langle{u}^p(t)\rangle_{{z}}\approx -\mathcal{K}\nabla_z\varrho$. One purpose of this paper is to carefully analyze how the model of \cite{rouse37} should be extended to more general cases, and the implications of the gradient-diffusion closure $\varrho\langle{u}^p(t)\rangle_{{z}}\approx -\mathcal{K}\nabla_z\varrho$. The other purpose is to consider the mechanisms governing $\langle{w}^p(t)\rangle_{{z}}$, which have been theoretically analyzed for HIT flows (e.g. \cite{maxey1987gravitational,tom19}), but have not been considered in detail for wall-bounded turbulent flows.

The equations governing $\varrho$ for arbitrary $St$ and $Fr$ are given by \eqref{VF} and \eqref{rhoeq}. Therefore, to understand and model the more general case, we must understand the role played by each term that appears in these equations. We will consider this, first considering their behavior in the quai-homogeneous regions, and then close to the wall where the flow is strongly inhomogeneous.

\subsection{Quasi-homogeneous region}\label{QHR}
In the quasi-homogeneous region away from the wall, the concentration profile is approximately constant ($\nabla_z\varrho \approx 0$) and from \eqref{VF} and \eqref{rhoeq} we obtain

\begin{align}
\varrho\langle{w}^p(t)\rangle_z=\Phi\approx\varrho\langle{u}^p(t)\rangle_z-\tau_p\varrho g,\label{rhoeqQH}
\end{align}
where $\langle{u}^p(t)\rangle_z$ is constant. In this regime, the mean particle momentum is governed by the Stokes terminal velocity $\tau_p g$, and a contribution from the average fluid velocity sampled by the particles, namely $\langle{u}^p(t)\rangle_z$. 

We may write
\begin{align}
\varrho\langle{u}^p(t)\rangle_z=\langle{u}(\bm{x},t)\delta (z^p(t)-z)\rangle,
\end{align}
where $\bm{x}^p(t)\bm{\cdot}\bm{e}_z=z^p(t)$ and $u(\bm{x},t)\equiv\bm{e}_z\bm{\cdot}\bm{u}(\bm{x},t)$. For particles that are (instantaneously) uniformly distributed in space, $\delta (z^p(t)-z)$ is constant, and so $\langle{u}(\bm{x},t)\delta (z^p(t)-z)\rangle=\langle{u}(\bm{x},t)\rangle\delta (z^p(t)-z)=0$. If the particles are non-uniform in space, $\langle{u}(\bm{x},t)\delta (z^p(t)-z)\rangle$ may be finite if there is a correlation between $u$ and $z^p(t)$. 

For the case where $\Phi< 0$, \citet{maxey1987gravitational} argued that the particles are preferentially swept around the downward moving side of vortices in the flow where $u<0$, leading to $\langle{u}^p(t)\rangle_z<0$. As a result, turbulence enhances the settling velocity of the particles compared to the Stokes settling velocity $\tau_p g$. In the regimes $St\ll 1$ and $St\gg 1$, $\langle{u}^p(t)\rangle_z=0$ because the particles are uniformly dstributed in these regimes. In the regime of rapidly settling particles, i.e. $Sv\equiv \tau_p g/\sqrt{\langle uu\rangle}\gg 1$, the correlation timescale of $u^p(t)$ vanishes \citep{bec14b,ireland16b} and as a result $\langle{u}^p(t)\rangle_z=0$ since there is no correlation between the particle motion and the local value of $u^p(t)$ in this limit. Similarly, in the regime $Sv\ll 1$, $\langle{u}^p(t)\rangle_z=0$ because the symmetry breaking effect of gravity that generates preferential sweeping vanishes in this regime.

For the case where $\Phi= 0$, then we must have $\langle{u}^p(t)\rangle_z=\tau_p g$. In this regime, the finitude of $\langle{u}^p(t)\rangle_z$ is due to the fact that in order for the vertical flux to be zero, the particles must preferentially sample upward moving regions of the fluid velocity field. Therefore, although particles moving down towards the wall may still experience the preferential sweeping mechanism that causes them to preferentially sample downward moving fluid, this contribution is overwhelmed by the contribution of particles moving up which necessarily experience strongly positive regions of the flow in order to satisfy $\Phi= 0$. 

In view of these considerations, we see that even for homogeneous flows, the importance of the preferential sweeping mechanism depends upon the boundary conditions in the system that determine the flux $\varrho\langle{w}^p(t)\rangle_z=\Phi$. The presence of the wall provides a way for the zero-flux scenario $\Phi=0$ to emerge, and was not considered in \citet{maxey1987gravitational} or \citet{tom19} where particle settling in an unbounded homogeneous flow was considered, for which the natural state that emerges is $\Phi<0$.

\subsection{Near wall region}\label{NWR}
As the particles approach the wall, gradients in the flow statistics become important and new mechanisms begin to control the particle settling velocity and concentration. In this case, all of the terms in \eqref{rhoeq}  are in principle important. 

The term R1 in \eqref{rhoeq} arises from the mean acceleration experienced by the particles due to gradients in their mean wall-normal velocity. This contribution vanishes for the zero-flux case $\Phi=0$ but is finite in general for $\Phi\neq 0$. The second term on the right hand side, R2 in \eqref{rhoeq}, describes a velocity arising from a diffusive flux. For fluid particles, the turbulent motion of the flow provides a mechanism for macroscopic diffusive transport (this will be discussed in more detail below). For particles with inertia, their velocity is partially decoupled from the local fluid velocity, and this decoupling introduces a second source of diffusion that is captured by $\tau_p S\nabla_z \varrho$. For $St\gg 1$, the PDF equation reduces to a Fokker-Planck equation and in this regime $\tau_p S\nabla_z \varrho$ is the sole source of diffusion \citep{bragg14b}. 

The third term, R3 in \eqref{rhoeq}, describes the turbophoretic drift velocity \citep{reeks83}. Physically, this drift velocity may be understood as follows: suppose the particles are moving in a region of the boundary layer where $\nabla_z\langle uu\rangle>0$. In this region, if the particle is moving towards the wall, then because they have come from regions where the flow has more TKE and because their response time is finite, they will be moving with greater kinetic energy than the local flow. On the other hand, if the particle is moving away from the wall then because they have come from regions where the flow has less TKE, they will be moving with less kinetic energy than the local flow. As a result, there is a symmetry breaking effect and the particles experience a velocity contribution towards the wall in regions where $\nabla_z\langle uu\rangle>0$, and the opposite in regions where $\nabla_z\langle uu\rangle<0$. In the limit $St\to 0$, the particle motion is governed only by the local flow, and so in this limit the turbophoretic effect vanishes. It also vanishes for $St\gg 1$ where the particles move ballistically through the boundary layer.

The fourth term, R4 in \eqref{rhoeq}, describes a source of momentum arising from preferential sampling of the local flow. In the previous section, this contribution was considered in the homogeneous region of the flow. In that region, $\langle{u}^p(t)\rangle_{{z}}$ can only be finite if $Sv$ is finite. Near the wall, however, $\langle{u}^p(t)\rangle_{{z}}$ can be finite even if $Sv=0$. This may be conceptually understood as follows. Suppose that due to the turbophoretic drift velocity and gravitational settling, the particles start to drift towards the wall, and their concentration builds up. If $\Phi=0$, the particles must escape the near wall region, and so they must preferentially sample regions of the flow where $u^p(t)>0$, leading to $\langle{u}^p(t)\rangle_{{z}}>0$. When gravity is also present, $\langle{u}^p(t)\rangle_{{z}}$ may be also affected by the preferential sweeping mechanism, however, this is likely to be a sub-leading effect unless $\Phi<0$, since if $\Phi=0$ we must have $\langle{u}^p(t)\rangle_{{z}}>0$ as discussed above (one exception is that for $St\gg1$, $\tau_p S\nabla_z \varrho$ is significant and provides a mechanism to remove particles from the wall, such that $\langle{u}^p(t)\rangle_{{z}}>0$ may not be required in order for particles to be able to escape the near-wall region; however, this is irrelevant since for $St\gg1$, $\langle{u}^p(t)\rangle_{{z}}=0$). As for the homogeneous region, $\langle{u}^p(t)\rangle_{{z}}=0$ for $St\to0$ and $St\to\infty$, the former because fluid particles uniformly sample the flow, and the latter because for $St\to\infty$ the particle motion is uncorrelated with the local fluid velocity.

Finally, the fifth (R5) and sixth (R6) terms on the rhs of \eqref{rhoeq} describe the Stokes settling velocity, and the diffusion induced velocity, respectively.

\subsection{Average fluid velocity seen by the particles}\label{AFS}

The average fluid velocity seen by the particles, $\langle{u}^p(t)\rangle_{{z}}$, plays in general an important role in determining the particle concentration and average vertical velocity. As noted earlier, the Rouse model for $\varrho$ effectively amounts to assuming an eddy viscosity, gradient-diffusion closure for this term. Here we consider this in more detail.

Analytical theories show that $\varrho\langle{u}^p(t)\rangle_{{z}}$ has the form \citep{reeks91,swailes97,reeks2012particle}
\begin{align}
\varrho\langle{u}^p(t)\rangle_{{z}}=\zeta\varrho-\sum_{n=1}^\infty\mathcal{D}^{[n]}\nabla_z^n\varrho,
\end{align}
where $\zeta$ is a drift coefficient, and $\mathcal{D}^{[n]}$ are diffusion coefficients that depend on $St,Sv$ and $z$ in general. The precise form of these coefficients is not quoted here since they depend upon the particular analytical theory used (see, e.g. \cite{bragg12b} for a detailed examination of the differences), and these details are not important for our discussion. In practice, in order to truncate this infinite expansion, most PDF based models of particle transport in turbulence assume that $u$ has Gaussian statistics, for which the series reduces exactly to \citep{reeks2012particle,swailes97,bragg12b}
\begin{align}
\varrho\langle{u}^p(t)\rangle_{{z}}=\zeta\varrho-\mathcal{D}^{[1]}\nabla_z\varrho.\label{uclose}
\end{align}
Most interestingly, however, the asymptotic analysis of \cite{sikovsky14} showed that for a wall-bounded flow, the regime $z^+\ll 1$ leads to $\sum_{n=1}^\infty\mathcal{D}^{[n]}\nabla_z^n\varrho\sim \mathcal{D}^{[1]}\nabla_z\varrho$ (where $+$ denotes that the variable has been normalized using wall units, in this case the friction length scale $\delta_\nu\equiv \sqrt{\nu/u_\tau}$, where $u_\tau$ is the wall friction velocity and $\nu$ is the fluid kinematic viscosity). This means that the contribution of the higher-order cumulants described by $\mathcal{D}^{[n]}$ for $n\geq 2$, which are neglected in \eqref{uclose} due to the Gaussian assumption, make a negligible contribution close to the wall. This is significant since it implies that \eqref{uclose} is accurate close to the wall, where the particle accumulation is strong and modeling is challenging. We also note that, as discussed in \cite{bragg12b}, models such as \cite{zaichik99} incorrectly set $\zeta=0$, which as we will soon see, has significant implications for modeling settling particles.

The fundamental difference between the drift $\zeta\varrho$ and diffusion $\mathcal{D}^{[1]}\nabla_z\varrho$ contributions in \eqref{uclose} is that whereas the diffusion contribution is only finite if there are finite gradients in the \emph{mean} particle distribution $\varrho$, the drift contribution may be finite even if $\nabla_z\varrho=0$, provided that there are inhomogeneities in the \emph{instantaneous} particle distribution and that those inhomogeneities are correlated with the local flow. For example, for fully-mixed fluid particles, the spatial distribution is uniform for all times, and so both the diffusion and drift contributions vanish, leading to $\varrho\langle{u}^p(t)\rangle_{{z}}=0$ \citep{bragg12b}. On the other hand, for settling inertial particles in a homogeneous flow, the diffusion contribution is zero, but the drift term is finite, capturing the preferential sweeping mechanism proposed by \citet{maxey1987gravitational}. Furthermore, an implication of the analysis in \cite{bragg12b}, that was subsequently demonstrated numerically in \cite{bragg12}, is that even in the absence of gravity, if the instantaneous distribution of the inertial particles in non-uniform, then $\zeta$ is also finite if the turbulence is inhomogeneous. Therefore, for setting inertial particles in wall-bounded turbulence, $\zeta$ may be finite both due to the preferential sweeping mechanism and also due to turbulence inhomogeneity. 

Phenomenological models that close $\varrho\langle{u}^p(t)\rangle_{{z}}$ using a gradient-diffusion hypothesis (such as the Rouse model) do not account for the drift contribution $\zeta\varrho$. The significance of this omission is that such models cannot account for Maxey's preferential sweeping mechanism (unless they account for it by modifying the Stokes settling velocity in the model). Note that for inhomogeneous flows, it is still the drift contribution $\zeta\varrho$ that formally accounts for preferential sweeping, and not the diffusive contribution. Therefore, this omission of $\zeta\varrho$ in gradient-diffusion closures is important for inhomogeneous flows, as well as homogeneous flows.

\subsection{Asymptotic behavior near the wall}\label{Asymp}
We now consider the asymptotic behavior for $z^+\ll 1$ for $\Phi=0$ first and then $\Phi\neq 0$. This asymptotic regime is of particular interest since it is in the near wall region that the particles are known to accumulate strongly in certain parameter regimes. Such an analysis has already been done for $g=0$ by \cite{sikovsky14}, and by \cite{johnson20} for the more restricted regimes $St\to 0, \Phi=0, g=0$. We want to explore how the behavior is modified when $g>0$ and $\Phi<0$.

For the analysis we will assume that $\varrho\langle{u}^p(t)\rangle_{{z}}$ is dominated by the diffusive contribution in \eqref{uclose} close to the wall where the gradients in $\varrho$ are largest, and we specify the diffusion coefficient $\mathcal{D}^{[1]}$ by its approximate form \citep{zaichik99}
\begin{align}
\mathcal{D}^{[1]}\approx \frac{\tau_L\langle uu\rangle}{1+\tau_p/\tau_L}.\label{Dform}
\end{align}
Furthermore, we will express all quantities in wall-units, denoted by a superscript $+$, which is accomplished by non-dimensionalizing the variables using the friction velocity $u_\tau$ and the friction length $\delta$. In the regime $z^+\ll1$, $\langle u^+u^+\rangle$ and $S^+$ have a power-law dependence on $z^+$ \citep{sikovsky14,johnson20} and $\tau_L$ is independent of $z^+$ \citep{kallio89}, according to which we may write for $z^+\ll1$
\begin{align}
\mathcal{D}^{{[1]}+}\sim a (z^+)^4,\label{Dnw}\\
\langle u^+u^+\rangle\sim b(z^+)^4,\label{UUnw}\\
S^+\sim c (z^+)^\gamma,\label{Snw}
\end{align}
where $a$ is defined through \eqref{Dform}, and $c$ depends in general upon $St^+$ and $Sv^+$. It is important to appreciate that while these forms are formally valid for $z^+\ll1$, we will later show that they apply for significantly larger values of $z^+$, as was also observed in \cite{sikovsky14}.

Combining \eqref{VF} and \eqref{rhoeq}, and using the diffusion approximation for $\varrho\langle{u}^p(t)\rangle_{{z}}$, we may write for $z^+\ll1$
\begin{align}
\Phi^+\sim -(St^+ c (z^+)^\gamma  +a (z^+)^4+\kappa^+ )\nabla^+\varrho^+-(c \gamma St^+(z^+)^{\gamma-1}+Sv^+)\varrho^+.\label{rhoeq_zsmall}
\end{align}
Here we have neglected the first and last terms in \eqref{rhoeq}, since as we will show later, our DNS data indicates that these terms are very small compared with the other terms in the equation. The solutions to \eqref{rhoeq_zsmall} have a number of different behaviors depending upon the regimes considered. We first consider the zero-flux case $\Phi^+=0$ and then consider $\Phi^+<0$.

For $\Phi^+=0$, if $St^+\ll 1$ then $\gamma=4+O(St^+)$ and the solutions to \eqref{rhoeq} are
\begin{align}
\varrho^+(z)\sim
\varrho^*\begin{cases}
(z^+)^{-4 c St^+/a}\exp\Big(\frac{Sv^+}{3 a(z^+)^{3}}\Big),\quad\text{for}\, a(z^+)^4\gg\kappa^+\\
\exp\Big(-\frac{c St^+(z^+)^4}{\kappa^+}-\frac{Sv^+z^+}{\kappa^+}\Big ),\quad\text{for}\, a(z^+)^4\ll\kappa^+\label{rhoStll1ZF}
\end{cases},
\end{align}
where $\varrho^*$ is an integration constant. This result shows that for $\kappa^+=Sv^+=0$, $\varrho^+$ grows as a power law $\varrho\sim \varrho^*(z^+)^{-4 c St^+/a}$ for $z^+\ll 1$, something already discussed in \citet{sikovsky14} and \citet{johnson20}. When $\kappa^+=0$ but $Sv^+>0$, the gravitational contribution causes $\varrho^+(z)$ to diverge exponentially fast as $z^+\to 0$. However, when $\kappa^+>0$, this regularizes the divergent behavior, causing $\varrho^+$ to asymptote to a constant as $z^+\to 0$.

When $St^+\geq O(1)$ and $\gamma<4$, three different behaviors for $\varrho^+$ emerge, depending on how the three contributions to the diffusion coefficient in \eqref{rhoeq_zsmall} compete. The three behaviors are
\begin{align}
\varrho^+(z)\sim
\varrho^*\begin{cases}
\exp\Big(\frac{St^+\gamma c(z^+)^{\gamma-4}}{a(4-\gamma)}    +\frac{Sv^+ }{3a^+ (z^+)^{3}}\Big),\quad\text{for}\, a(z^+)^4\gg St^+ c (z^+)^\gamma+  \kappa^+\\
(z^+)^{-\gamma}\exp\Big(\frac{Sv^+ (z^+)^{1-\gamma}}{St^+ c(\gamma-1)}\Big),\quad\text{for}\, St^+ c (z^+)^\gamma\gg a(z^+)^4 +  \kappa^+\\
\exp\Big(-\frac{St^+ c (z^+)^\gamma}{\kappa} -\frac{Sv^+ z^+}{\kappa^+}\Big),\quad\text{for}\, \kappa^+\gg St^+ c (z^+)^\gamma+ a(z^+)^4 \label{rhoStgeq1ZF}
\end{cases}.
\end{align}
For $\kappa^+=Sv^+=0$, the second regime gives the power-law solution $\varrho^+(z)\sim \varrho^*(z^+)^{-\gamma}$ and $\varrho^+ S^+\sim$ constant, as described in \cite{sikovsky14}. However, the first regime gives $\varrho^+(z)\sim \varrho^*\exp(St^+\gamma c(z^+)^{\gamma-4}/a(4-\gamma))$ which grows exponentially as $z^+$ decreases since for $St^+\geq O(1)$, $\gamma<4$. This behavior does not seem to have been considered in \cite{sikovsky14}, and this is most likely because for $St^+\geq O(1)$, $\gamma<4$ so that the regime $a(z^+)^4\gg St^+ c (z^+)^\gamma+\kappa^+$ is not accessible in the limit $z^+\to 0$ that was considered in \cite{sikovsky14}. However, since the near-wall scaling in \eqref{Dnw}-\eqref{Snw} actually applies in practice for $z^+\leq O(1)$, then this regime may be accessible and important. One significant difference between the asymptotic behaviors $\varrho^+(z)\sim \varrho^*(z^+)^{-\gamma}$ and $\varrho^+(z)\sim \varrho^*\exp(St^+\gamma c(z^+)^{\gamma-4}/a(4-\gamma))$ is that the former grows more slowly with decreasing $z^+$ as $\gamma$ decreases, while the latter grows faster with decreasing $z^+$ as $\gamma$ decreases. We will return to this point when discussing our DNS results.

For $\kappa^+=0$ but $Sv^+> 0$, the behavior depends on $\gamma\in[0,4]$, the upper limit corresponding to $St^+= 0$, while $\gamma=0$ corresponds to the ballistic regime that occurs for $St^+\to \infty$. If $\gamma>1$, then gravitational settling will dominate the growth of $\varrho^+$ as $z^+$ decreases, but the growth differs for regimes 1 and 2 in \eqref{rhoStgeq1ZF}, the growth in regime 1 being the fastest since $\gamma<4$ for $St^+\geq O(1)$. For $\gamma<1$, we recover the power-law solution $\varrho^+(z)\sim \varrho^*(z^+)^{-\gamma}$ in regime 2, while for regime 1 we recover $\varrho^+(z)\sim \varrho^*\exp(St^+\gamma c(z^+)^{\gamma-4}/a(4-\gamma))$. These behaviors emerge since for $\gamma<1$, the turbophoretic velocity dominates over the Stokes settling velocity in the regime $z^+\ll 1$. Interestingly, however, in analogy with the effect of gravity on the relative motion of inertial particle-pairs in turbulence \citep{ireland16b}, when $Sv^+$ increases, the timescale of the fluid seen by the particle reduces, and the particles fall rapidly through the flow in the regime $Sv^+\gg 1$. As a result of this, the path-history effect of particle inertia reduces, and hence $\gamma$ increases (since the particle velocities are more closely linked to the behavior of the local fluid velocity field). Therefore, the effect of gravity on the dependence of $\varrho^+$ with $z^+$ is also through its implicit effect on the exponent $\gamma$. Finally, for $\kappa^+>0$, $\varrho^+(z)$ approaches a constant for $z^+\to 0$, reflecting the fact that $\kappa^+>0$ regularizes the divergent behavior since it provides a finite source of diffusion as $z^+\to0$.

Due to \eqref{VF}, if $\kappa^+=0$, $\langle w^p(t)\rangle^+_z=0$ when $\Phi^+=0$, but if $\kappa^+>0$ then using the previous results for $\varrho^+$ together with \eqref{VF} we obtain for $St^+\ll1$
\begin{align}
\langle w^p(t)\rangle^+_z\sim
\begin{cases}
-\kappa^+(12 c St^+(z^+)^{-1}+3 Sv^+(z^+)^{-4})/3 a,\quad\text{for}\, a(z^+)^4\gg\kappa^+\\
-4 c St^+ z^3-Sv^+,\quad\text{for}\, a(z^+)^4\ll\kappa^+\\
\end{cases},\label{wav_zF}
\end{align}
while for $St^+\geq O(1)$ we obtain
\begin{align}
\langle w^p(t)\rangle^+_z\sim
\begin{cases}
-\kappa^+(St^+\gamma c (z^+)^{\gamma-5}+Sv^+ (z^+)^{-4})/a,\quad\text{for}\, a(z^+)^4\gg St^+ c (z^+)^\gamma+  \kappa^+\\
-\kappa^+(\gamma (z^+)^{-1}+(Sv^+/St^+ c)(z^+)^{-\gamma}),\quad\text{for}\, St^+ c (z^+)^\gamma\gg a(z^+)^4 +  \kappa^+\\
-\gamma St^+ c (z^+)^{\gamma-1}-Sv^+,\quad\text{for}\, \kappa^+\gg St^+ c (z^+)^\gamma+ a(z^+)^4 \label{wStgeq1ZF}
\end{cases}.
\end{align}
This shows that for $z^+\to 0$ we always have $\langle w^p(t)\rangle^+_z\to -Sv^+$, unless $\gamma<1$, for which $\langle w^p(t)\rangle_z$ actually diverges as $z^+\to 0$. Although this behavior may seem surprising, it occurs because in regime 3 of \eqref{rhoStgeq1ZF}, $\nabla^+\varrho^+$ (but not $\varrho^+$) diverges as $z^+\to0$ and hence so also must $\langle w^p(t)\rangle^+_z$ in order to satisfy \eqref{VF}.

We now consider $\Phi^+< 0$, for which results may be derived from \eqref{rhoeq_zsmall}, although the analysis is now considerably more involved and analytical solutions cannot be derived for all regimes. If we take $\kappa^+=0$ (as will be done in our DNS for this case) then we obtain for $St^+\ll1$ 

\begin{align}
\varrho^+(z)\sim
\begin{cases}
\Phi^+ (z^+)^{-3}/4 a +\varrho^* (z^+)^{-4 c St^+/a},\quad\text{for}\, 4 c St^+ (z^+)^3\gg Sv^+\\
-\Phi^+/Sv^+ +\varrho^*\exp\Big(\frac{Sv^+}{3 a(z^+)^{3}}\Big),\quad\text{for}\, 4 c St^+ (z^+)^3\ll Sv^+
\end{cases},\label{rhoStll1CF}
\end{align}
and for $St\geq O(1)$
\begin{align}
\varrho^+(z)\sim
\begin{cases}
-\Phi^+ (z^+)^{1-\gamma}/St^+ c +\varrho^* (z^+)^{-\gamma},\quad\text{for}\, \gamma St^+ c(z^+)^{\gamma-1}\gg Sv^+\\
 -\Phi^+/Sv^+ +\varrho^*\exp\Big(\frac{Sv^+ (z^+)^{1-\gamma}}{St^+ c(\gamma-1)}\Big),\quad\text{for}\,\gamma St^+ c(z^+)^{\gamma-1}\ll Sv^+\\
\end{cases},\label{rhoStgeq1CF}
\end{align}
where we have split up the regimes into those where either the turbophoretic velocity is much larger than $Sv^+$, or vice-versa, since the solution for the overlap region is very involved. Furthermore, \eqref{rhoStgeq1CF} only applies for $St^+ c (z^+)^\gamma\gg a(z^+)^4 $. There is no analytical solution for the regime $St^+ c (z^+)^\gamma\ll a(z^+)^4 $ when $\Phi^+$ is finite except for very restricted cases. 

For regime 1 of \eqref{rhoStll1CF}, the contribution involving $\Phi^+$ is negative for $\Phi^+<0$, and therefore the second term must always be dominant in order to preserved the requirement $\varrho^+\geq 0$. For the other regimes in \eqref{rhoStll1CF} and \eqref{rhoStgeq1CF}, the contribution involving $\Phi^+$ is always positive for $\Phi^+<0$, however, this contribution is negligible in the limit $z^+\to 0$. In the absence of turbulence, $\langle w^p(t)\rangle^+_z=-Sv^+$, and $\varrho^+=-\Phi^+/Sv^+$. For $z^+\to 0$, $u^+\to 0$ and so we might expect $\varrho^+=-\Phi^+/Sv^+$ to be recovered in this limit (especially for $St^+\ll1$), but the solutions in \eqref{rhoStgeq1CF} show otherwise. The reason for this is that \eqref{rhoeq_zsmall}  is singularly perturbed with respect to the magnitude of the turbulent fluctuations. As such, the diffusion term in \eqref{rhoeq_zsmall} which involves $\nabla_z\varrho$ remains finite even though the diffusion coefficient tends to zero (when $\kappa^+=0$) for $z^+\to 0$, and for this reason $\varrho^+=-\Phi^+/Sv^+$ is not recovered in this limit in a turbulent flow. Finally, for the case $\Phi^+<0$ with $\kappa^+=0$, we obtain $\langle w^p(t)\rangle^+_z=\Phi^+/\varrho^+$ from \eqref{VF}, and hence the asymptotic solutions for $\langle w^p(t)\rangle^+_z$ are readily obtained from \eqref{rhoStll1CF} and \eqref{rhoStgeq1CF}.


\section{Direct Numerical Simulations}\label{sec:DNS}

\subsection{Equations of motion}\label{subsec:Carrier}
In order to explore the role of each term appearing in \eqref{VF} and \eqref{rhoeq}, and to test the asymptotic results in \S\ref{Asymp}, we use data from DNS of settling inertial particles in a horizontal, fully developed, incompressible turbulent open channel flow. The DNS solves the incompressible Navier-Stokes equations
\begin{equation}
\partial_t \bm{U}+ (\bm{U\cdot\nabla_x})\bm{U}=-\frac{1}{\rho_f}\bm{\nabla_x}p+\nu \bm{\nabla_x}^2\bm{U},\label{eq:method_momen}
\end{equation} 
where $\bm{U}(\bm{x},t)$ is the fluid velocity, $p(\bm{x},t)$ is the pressure, $\nu$ is the fluid kinematic viscosity, and $\rho_{f}$ is the fluid density. A pseudospectral method is employed in the periodic directions (streamwise $x$ and spanwise $y$), and second-order finite differences are used for spatial discretization in the wall-normal, $z$ direction. The solution is advanced in time using a third-order Runge-Kutta (RK3) scheme. The incompressibility constraint $\bm{\nabla_x\cdot U}=0$ is satisfied by prescribing the pressure via the solution of its Poisson equation $\bm{\nabla_x}^2p=\bm{\nabla_x U:\nabla_x U}$ \citep{pope}.

The flow field from DNS has been tested and validated by comparison with published data in multiple configurations; e.g., planar Couette flow at $\Rey_\tau=40$ \citep{wang2019modulation}, wall-bounded channel flow at $\Rey_\tau=227,630$ \citep{wang2019inertial}, and open channel flow at $\Rey_\tau=200,550,950$ \citep{wang_richter_2019}. The carrier phase velocity fields from the DNS are used to integrate the trajectories of inertial particles using the standard point-particle approach.

In the DNS, inertial particles are tracked by solving \eqref{eomx} and \eqref{eomv}, using the solution to \eqref{eq:method_momen} to obtain $\bm{U}(\bm{x},t)$, which is interpolated to the particle position using a sixth-order Lagrange method to obtain $\bm{U}^p(t)$. The particle Reynolds number appearing in $\Psi$ is defined as $Re_p\equiv \|\bm{U}^p(t)-\bm{v}^p(t) \| d_p / \nu $, which is based on the magnitude of the particle slip velocity $\|\bm{U}^p(t)-\bm{v}^p(t)\|$ and the particle diameter $d_{p}$. In this work, the average $Re_p$ is less than $1$, which is far smaller than the suggested maximum $Re_{p} \approx 800$ for the Schiller-Naumann \citep{schiller1933ber} model. As a result of the low $Re_{p}$, the correction to the Stokes drag is minimal in this study. 

Our DNS code has been validated in \citet{wang2019inertial} for inertial particles in the range $St^+=30-2000$ by comparisons against the code of \citet{capecelatro2013euler} as well as the experimental results of \cite{fong2019velocity}.\\

\subsection{Boundary conditions and numerical parameters}\label{subsec:BCs}

\begin{figure}
\centering
\includegraphics[width=0.9\textwidth]{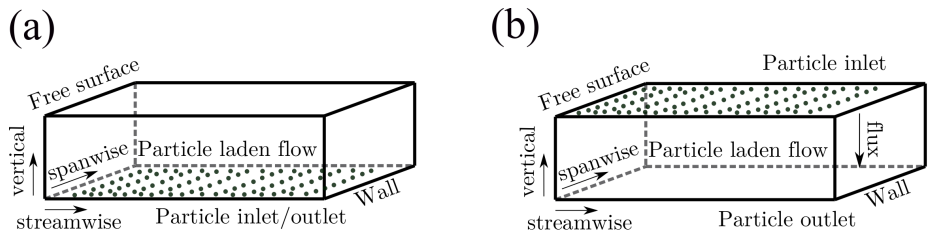}
\caption{(a) zero-flux configuration, in which a constant particle concentration is maintained using a reservoir just beneath the wall; (b) constant flux configuration, in which particles are initialized at the top of the domain and removed/replaced when they reach the wall.}
\label{fig:configuration}
\end{figure}

We solve \eqref{eq:method_momen} at $Re_{\tau} = 315$, using a constant pressure gradient to force the flow. The streamwise $x$ and spanwise $y$ directions are periodic, and the wall at $z = 0$ imposes a no-slip condition on the fluid velocity field. At the upper wall, $z = H$, a free-slip (i.e., zero-stress) condition is imposed on the fluid velocity. This setup provides a canonical case of wall-bounded turbulence, within which two distinct particle configurations are considered; figure \ref{fig:configuration} provides schematics of these two configurations.

In the first configuration, termed the ``zero-flux'' case (indicated by $ZF$ and designed to provide $\Phi = 0$), a constant number concentration of particles is maintained in a reservoir just beneath the wall at $z = 0$, which serves as Dirichlet condition on the particle concentration in the Lagrangian framework. At the upper domain boundary $z = H$, the particles rebound elastically, which is equivalent to a no-flux condition on the particle concentration. The Brownian diffusion term included in \eqref{eomx} enables the particles to be suspended into the flow, and we set $\kappa=\nu$. In the interior of the domain, this diffusive contribution is negligible compared to the particle motion produced by the velocity field, and only dominates near the wall. In this first no-flux configuration, a statistically steady-state particle distribution is established after a sufficient time. This zero-flux configuration is the same as that used in \citet{RichterBLM2018}.

In the second particle configuration, termed the ``constant flux'' case (denoted by $CF$ and designed to provide $\Phi < 0$), the Lagrangian particles are instead placed at the upper boundary ($z = H$) at a random location on the $x-y$ plane and given an initial vertical velocity equal to their terminal Stokes settling velocity $\tau_p g$ (the other two particle velocity components are set to zero). From here, the particles settle by gravity through the system and when they reach the wall they are removed. For each particle removed at the wall, one is re-introduced at the upper boundary at a random location, and therefore the total number of particles $N_{p}$ is constant throughout the simulation (in contrast to the first no-flux configuration). After a sufficient time, the concentration profile and vertical flux attain statistical stationarity, with the net flux $\Phi$ independent of $z$ and having magnitude that varies with $\tau_p$. Note that for this constant-flux configuration, the diffusion term in \eqref{eomx} is not used, unlike the zero-flux configuration where it is required to enable particle suspension into the flow from the wall.

For each of the two configurations, six different simulations are performed, where the particle Stokes number is systematically increased. These are presented in table \ref{tab:Table_1}, where case number 0 refers to the lowest Stokes number and 5 refers to the highest. We define two different Stokes numbers: $St^{+}$, which is based on viscous wall units and ranges between $0.003$ and $46.5$, and $St_{K}$, which is based on the vertically-averaged Kolmogorov timescale $\tau_{K}$ in the flow and ranges between $3.4 \times 10^{-4}$ to $5.1$. 

In order to isolate the effect of particle inertia on settling through wall-bounded turbulence, the gravitational acceleration $g$ is varied with each case in order to maintain a constant settling parameter $Sv^+\equiv\tau_p g/u_{\tau} = 2.5 \times 10^{-2}$, where $u_{\tau}$ is the friction velocity of the flow. This corresponds to $Sv_K\equiv\tau_p g/u_{K} = 7.4 \times 10^{-4}$ when the Stokes settling velocity is normalized by the vertically-averaged Kolmogorov velocity. We choose this value since it was also used in \citet{RichterBLM2018} allowing us to compare our results to theirs, and to further understand how the modified Rouse model that they considered, which was found to be accurate for $St\ll1$, must be modified for predicting the case of general $St$. Moreover, since $Sv^+$ is fixed in our simulations, any observed changes in the particle statistics are solely due to changes in the Stokes number, not the settling number, and this aids in understanding the results. In the environment where $g$ is constant, both $St^+$ and $Sv^+$ change as $\tau_p$ is varied, and this case will be considered in future work.

\begin{table}
\centering

\begin{tabular}{cclclcccc}
Case &     &       & $St^+$ & $St_K$ & $Sv^+$ & $Sv_K$    \\
$ZF$ & $CF$  &              &        &        &       &        \\
0   & 0    &   & 0.003  & 3.4e-4 & $2.5 \times 10^{-2}$ & $7.4 \times 10^{-4}$    \\
1   & 1    &       & 0.93   & 0.102  & $2.5 \times 10^{-2}$ & $7.4 \times 10^{-4}$ \\
2   & 2    &      & 2.79   & 0.306  & $2.5 \times 10^{-2}$ & $7.4 \times 10^{-4}$ \\
3   & 3    &        & 4.65   & 0.51   & $2.5 \times 10^{-2}$ & $7.4 \times 10^{-4}$ \\
4   & 4    &       & 9.30   & 1.02   & $2.5 \times 10^{-2}$ & $7.4 \times 10^{-4}$ \\
5   & 5    &       & 46.5   & 5.10   & $2.5 \times 10^{-2}$ & $7.4 \times 10^{-4}$ \\
\hline
\end{tabular}

\caption{Summary of the simulations. All cases are turbulent open channel flow at $Re_{\tau} = 315$. $ZF$ refers to the ``zero-flux" case while $CF$ refers to the ``constant flux'' case. Case numbers 0 -- 5 indicate increasing Stokes numbers.}
\label{tab:Table_1}

\end{table}

\section{Results \& Discussion}\label{sec:Results}

\subsection{Behavior of average particle distribution and vertical velocity}

\begin{figure}
\centering
\begin{overpic}
[width=0.99\textwidth,trim={0mm 0mm 0mm 0mm}, clip]{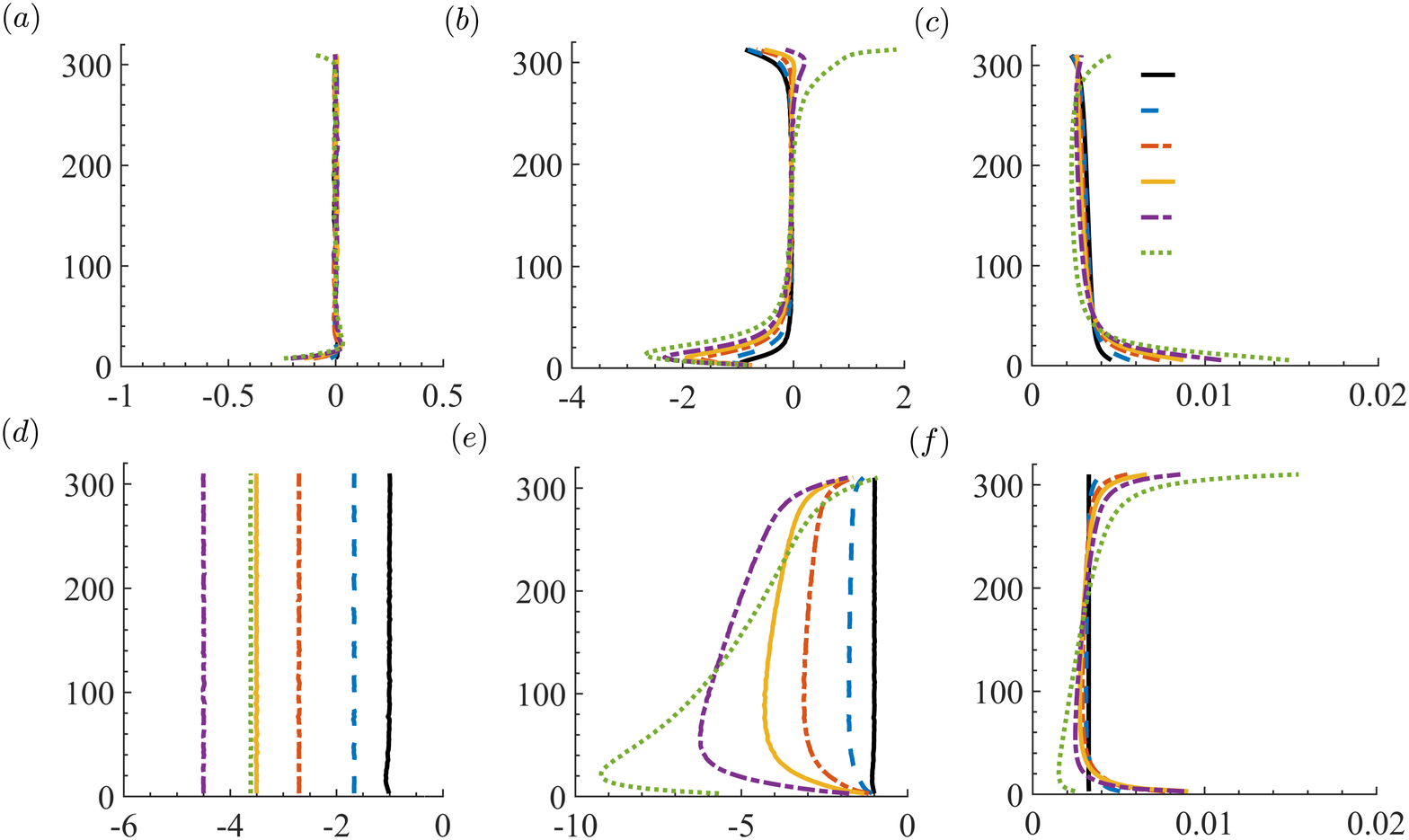}
\put(322,203){\scriptsize{{Case 0}}}
\put(322,194){\scriptsize{{Case 1}}}
\put(322,185){\scriptsize{{Case 2}}}
\put(322,175){\scriptsize{{Case 3}}}
\put(322,166){\scriptsize{{Case 4}}}
\put(322,156){\scriptsize{{Case 5}}}
\put(176,-6){$\langle w^p(t)\rangle_z/\tau_p g$}
\put(322,-6){$\varrho$}
\put(75,-6){$\Phi$}
\put(8,165){\rotatebox{90}{$z^+$}}
\put(130,165){\rotatebox{90}{$z^+$}}
\put(252,165){\rotatebox{90}{$z^+$}}
\put(8,55){\rotatebox{90}{$z^+$}}
\put(130,55){\rotatebox{90}{$z^+$}}
\put(252,55){\rotatebox{90}{$z^+$}}
\end{overpic}
\caption{DNS results for the total mass flux $(a,d)$, normalized average vertical particle velocity $(b,e)$, and spatial distribution $(c,f)$. Different colors correspond to the different cases. Plots (a,b,c) and (d,e,f) correspond to the zero-flux and constant-flux cases, respectively.}
\label{fig:terms}
\end{figure}

In figure~\ref{fig:terms} we show results for the average total mass flux $\Phi$, normalized vertical velocity $\langle w^p(t)\rangle_z/\tau_p g$, and spatial distribution $\varrho$ for each of the cases, and for both the zero-flux (a,b,c) and constant-flux (d,e,f) configurations. We begin by describing the results for $\Phi, \langle w^p(t)\rangle_z, \varrho$, and will then turn to examine the underlying cause of their behavior in terms of the various mechanisms described by \eqref{VF} and \eqref{rhoeq}. 

The results in figure~\ref{fig:terms} for $\Phi$ for the zero-flux case are computed using \eqref{VF}, and the small deviations of $\Phi$ from zero near the wall are due to statistical and numerical error when differentiating the DNS data for $\varrho$. For the zero-flux configuration, $\langle w^p(t)\rangle_z$ is zero away from the boundaries, but takes on finite values near the boundaries due to the contribution to the particle motion from the diffusive term involving $\kappa$ in \eqref{eomx}. As the wall is approached, $\varrho$ begins to increase significantly, indicating that the particles accumulate near the wall, and for the cases considered, max($\varrho$) increases monotonically with increasing $St$. 

For the constant-flux configuration, $\Phi$ varies non-monotonically with $St$, and is maximum for Case 4. As we will discuss momentarily, this non-monotonic behavior is due to turbulence since in the absence of turbulence, $\Phi$ would be independent of $St$ because $Sv$ is held constant in our DNS. Clearly, turbulence strongly influences this vertical mass flux, leading to enhancements of up to a factor of 4.5 for the cases considered. The average vertical velocity $\langle w^p(t)\rangle_z$ increases at all heights with increasing $St^+$, except in going from Case 4 to 5 where $\langle w^p(t)\rangle_z$ reduces with increasing $St^+$ in the upper portion of the domain. The results show that for Cases 1-3 as the particles move from the upper boundary towards the wall, they pass through a significant region where $\langle w^p(t)\rangle_z$ only slightly increases. As they get close to the wall, however, $\langle w^p(t)\rangle_z/\tau_p g$ suddenly drops due to the fluid velocity fluctuations reducing as the wall is approached. For Cases 4 and 5, $\langle w^p(t)\rangle_z$ varies significantly with $z^+$ throughout the entire domain, increasing significantly as $z^+$ is reduced down to $z^+\approx 20$, below which $\langle w^p(t)\rangle_z$ reduces significantly. We note that for all cases, $\langle w^p(t)\rangle_z/\tau_p g$ drops as the wall is approached, but never actually reaches unity, despite the fact that the turbulent fluctuations vanish as $z^+\to 0$. 

It is important to emphasize that since $Sv^+= 2.5 \times 10^{-2}$ in our DNS (fixed for all $St^+$), the actual settling velocities of the particles are small compared with the velocity scales in the flow, i.e. $\langle w^p(t)\rangle_z\ll u_\tau\,\forall z$. Nevertheless, relative to $\tau_p g$, the enhancement to the particle settling velocity due to turbulence is significant, with $\langle w^p(t)\rangle_z/\tau_p g$ attaining values up to almost 10. In many studies of settling enhancement due to turbulence, the enhancement is quantified by comparison with some fluid velocity, e.g. $(\langle w^p(t)\rangle_z+\tau_p g)/u_\tau$ (or compared with $u_K$ for HIT, e.g. \cite{wang1993settling,ireland16b}), which will always be $\ll1$ for $Sv^+\ll 1$ and therefore essentially disguising any significant enhancements relative to $\tau_p g$ that may occur for $Sv^+\ll 1$. Instead, one ought to compare $\langle w^p(t)\rangle_z$ with $\tau_p g$, not $u_\tau$, and when we do this we find that turbulence enhances the settling velocity of the particles relative to $\tau_p g$ significantly, even if $Sv^+\ll 1$.

Concerning $\varrho$, for the constant-flux case we observe that away from the upper boundary, $\varrho$ decreases slightly as $z^+$ decreases, until close to the wall where it sharply increases, indicating a near-wall accumulation of the particles. For this constant-flux case where there is no diffusion and \eqref{VF} reduces to $\varrho=\Phi/\langle w^p(t)\rangle_z$, $\varrho$ necessarily increases close to the wall if $\langle w^p(t)\rangle_z$ decreases as $z^+$ decreases, w we would expect when turbulence plays a role in the particle motion.

\FloatBarrier

\begin{figure}
\centering
\begin{overpic}
[width=0.9\textwidth,trim={0mm 0mm 0mm 0mm}, clip]{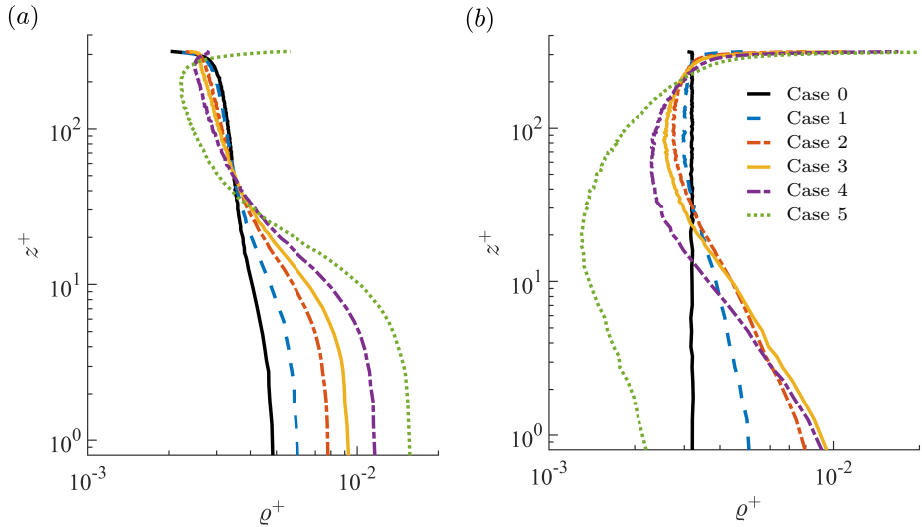}
\put(294,160){\scriptsize{{Case 0}}}
\put(294,151){\scriptsize{{Case 1}}}
\put(294,142){\scriptsize{{Case 2}}}
\put(294,133){\scriptsize{{Case 3}}}
\put(294,124){\scriptsize{{Case 4}}}
\put(294,115){\scriptsize{{Case 5}}}
\end{overpic}
				
\caption{Results for $\varrho^+$ plotted in a log-log scale to emphasize the behavior close to the wall. (a) zero-flux configuration, (b) constant flux configuration.}
\label{fig:Rholog}
\end{figure}

In figure \ref{fig:Rholog} we plot the results for $\varrho^+$ in a log-log scale in order to examine the behavior close to the wall. For the zero-flux configuration, we find that for $10\lesssim z^+\lesssim 30$, $\varrho^+$ increases as a power-law, as observed in \citet{sikovsky14} and \citet{johnson20} and discussed in \S\ref{Asymp}. However, our results indicate that the power-law exponent is increasing with increasing $St^+$. This is not consistent with the power-law of regime 2 in \eqref{rhoStgeq1ZF}, namely $\varrho^+\sim\varrho^* (z^+)^{-\gamma}$, which is the power-law behavior discussed in \citet{sikovsky14} and \citet{johnson20}, since $\gamma$ will monotonically decrease with increasing $St^+$. Since the power-law we observe occurs where $z^+>1$ then we may expand regime 1 of \eqref{rhoStgeq1ZF} and obtain
\begin{align}
\begin{split}
 \varrho^+&\sim\varrho^*   \exp\Big(\frac{St^+\gamma c(z^+)^{\gamma-4}}{a(4-\gamma)}    +\frac{Sv^+ }{3a^+ (z^+)^{3}}\Big)\\
 &\sim\varrho^*\Big(1+\frac{St^+\gamma c(z^+)^{\gamma-4}}{a(4-\gamma)}\Big)+O((z^+)^{2(\gamma-4)}),
\end{split}
\end{align}
which is valid for $\gamma>1$. This shows that as $\gamma$ decreases from 4 with increasing $St^+$, the power-law exponent describing $\varrho^+$ (namely $\gamma-4$) grows, which is consistent with the behavior observed in \ref{fig:Rholog} for the zero-flux case in the regime $10\lesssim z^+\lesssim 30$. For $z^+<10$, the results show that $\varrho^+$ becomes constant due to the contribution from diffusion in \eqref{eomx}, and as predicted by \eqref{rhoStll1ZF} and \eqref{rhoStgeq1ZF}.

For the constant flux configuration, the results in figure \ref{fig:Rholog} show that for $z^+\leq O(10)$, $\varrho^+$ increases as a power-law as the wall is approached. In going from Case 4 to 5, the power-law exponent reduces, indicating that for these cases, the behavior corresponds to the regime \eqref{rhoStgeq1CF} where the role of gravity is subdominant. For the other cases, the results indicate that the power-law exponent for $\varrho^+$ is increasing with increasing $St^+$. This is consistent with the prediction in \eqref{rhoStll1CF} for the regime where the role of gravity is subdominant, however, this is only supposed to be valid for $St^+\ll1$, and therefore cannot explain the behavior for all of the remaining cases. The increase of the power-law exponent for $\varrho^+$ with increasing $St^+$ is not consistent with the prediction in \eqref{rhoStgeq1CF} for the regime where the role of gravity is subdominant. The behavior must therefore correspond to that which emerges in the regime where $St^+ c (z^+)^\gamma\ll a(z^+)^4 $, for which we cannot derive an analytical prediction for $\varrho^+$ as discussed in \S\ref{Asymp}. 

The results in \eqref{rhoStll1CF} and \eqref{rhoStgeq1CF} also predict that for $\Phi^+<0$, then provided $\gamma>1$, $\varrho^+$ will grow at an exponential rate as $z^+$ is decreased, for sufficiently small $z^+$. We do not observe evidence of this from the DNS, however. The simplest explanation is simply that we do not have data at sufficiently small $z^+$ in order to be able to observe this. Another explanation could be that this near-wall, exponential accumulation of particles takes a significant amount of time to develop, and that our DNS has not been run for long enough in order for the very near wall accumulation to reach a steady state. Future investigations are required to clarify this issue.

\begin{figure}
\centering
\begin{overpic}
[width=0.9\textwidth,trim={0mm 0mm 0mm 0mm}, clip]{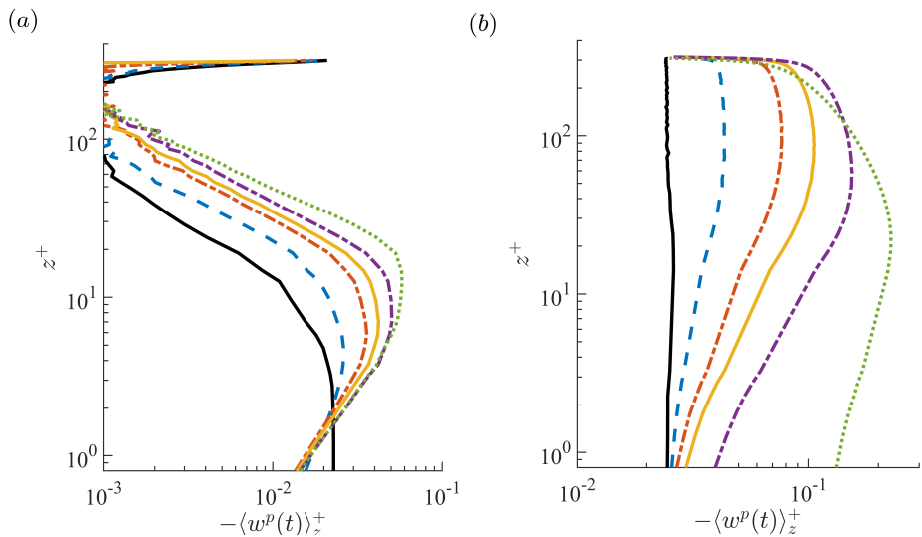}
\end{overpic}
				
\caption{Results for $-\langle w^p(t)\rangle^+_z$ plotted in a log-log scale to emphasize the behavior close to the wall. (a) Zero-flux configuration, (b) constant flux configuration. Legend is the same as figure \ref{fig:Rholog}.}
\label{fig:wlog}
\end{figure}

In figure \ref{fig:wlog} we likewise plot the results for $-\langle w^p(t)\rangle^+_z$ in a log-log scale in order to examine the behavior close to the wall. For the constant flux case, figure \ref{fig:wlog}(b), $\langle w^p(t)\rangle^+_z$ decays as a power law for $z^+\leq O(10)$, which must be the case since $\varrho^+$ increases as a power-law in this region, and $\varrho^+\langle w^p(t)\rangle^+_z=\Phi^+$ is a constant. The results for the zero-flux case are, however, unexpected. For Case 1,  $\langle w^p(t)\rangle^+_z$ asymptotes to one, as expected based on \eqref{wav_zF}. However, for Cases 2-5, the results show that  $\langle w^p(t)\rangle^+_z$ decays as a power-law in $z^+$ near the wall, but with $\langle w^p(t)\rangle^+_z<Sv^+$. This does not agree with \eqref{wav_zF} that predicts that while $\langle w^p(t)\rangle^+_z$ decays as a power-law in $z^+$, it should asymptotically approach $Sv^+$ from above, i.e. $\langle w^p(t)\rangle^+_z\geq Sv^+$. At present we are not able to explain this discrepancy, but future DNS using more particles that are better able to resolve the very near wall particle motion will help in examining the asymptotic behavior more fully.

\begin{figure}
\centering
	\subfloat{}
\begin{overpic}
[width=0.9\textwidth,trim={0mm 0mm 0mm 0mm}, clip]{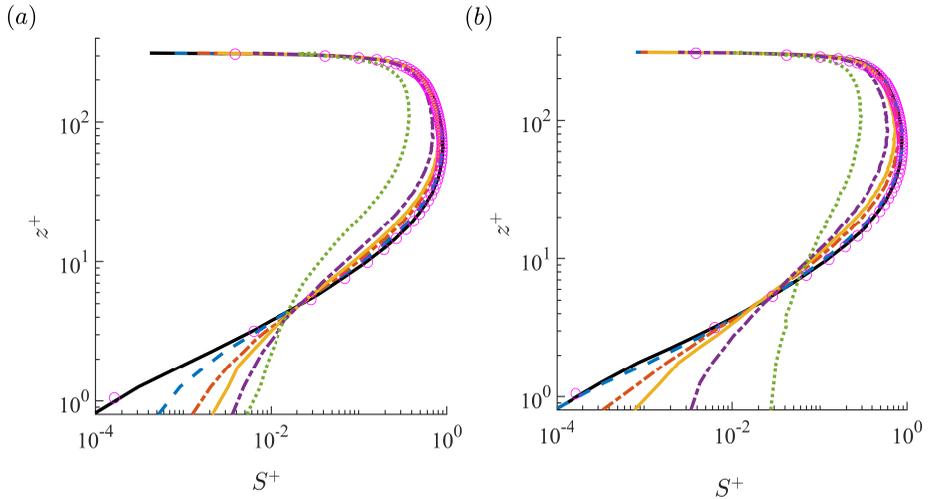}
\end{overpic}

				
\caption{Results for $S^+$ plotted in a log-log scale to emphasize the behavior close to the wall. (a) Zero-flux configuration, (b) constant flux configuration. Legend is the same as figure \ref{fig:Rholog}, except for $\circ$ which corresponds to the fluid wall-normal Reynolds stress $\langle u^+ u^+\rangle$.}
\label{fig:Slog}
\end{figure}

In deriving the asymptotic predictions for $\varrho^+$ in \S\ref{Asymp}, we assumed, following previous analysis, that $S^+\propto (z^+)^\gamma$. In figure  \ref{fig:Slog} we plot $S^+$, and the results show that close to the wall, a power-law behavior does in fact emerge, just as previously observed for the case without gravity \citep{sikovsky14,johnson20}. Our data does not allow us to explore $S^+$ at small enough $z^+$ to reliably measure the exponent $\gamma$ which would be required for a quantitative test of the asymptotic predictions of \S\ref{Asymp}. 

Summarizing, several features of the results near the wall are described well by the asymptotic results in \S\ref{Asymp}, while others do not agree. Future work should involve running the DNS for longer and with more particles in order to improve the particle statistics at $z^+\ll 1$, and to see whether these discrepancies are resolved at sufficiently small $z^+$ (recalling that the asymptotic results are strictly for $z^+\ll1$). Moreover, such data would enable $\gamma$ to be measured, allowing for a more quantitative test of the asymptotic predictions.

\FloatBarrier

\FloatBarrier

\subsection{Mechanisms controlling the wall-normal particle motion}

\begin{figure}
\centering
\begin{overpic}
[width=0.99\textwidth,trim={0mm 0mm 0mm 0mm}]{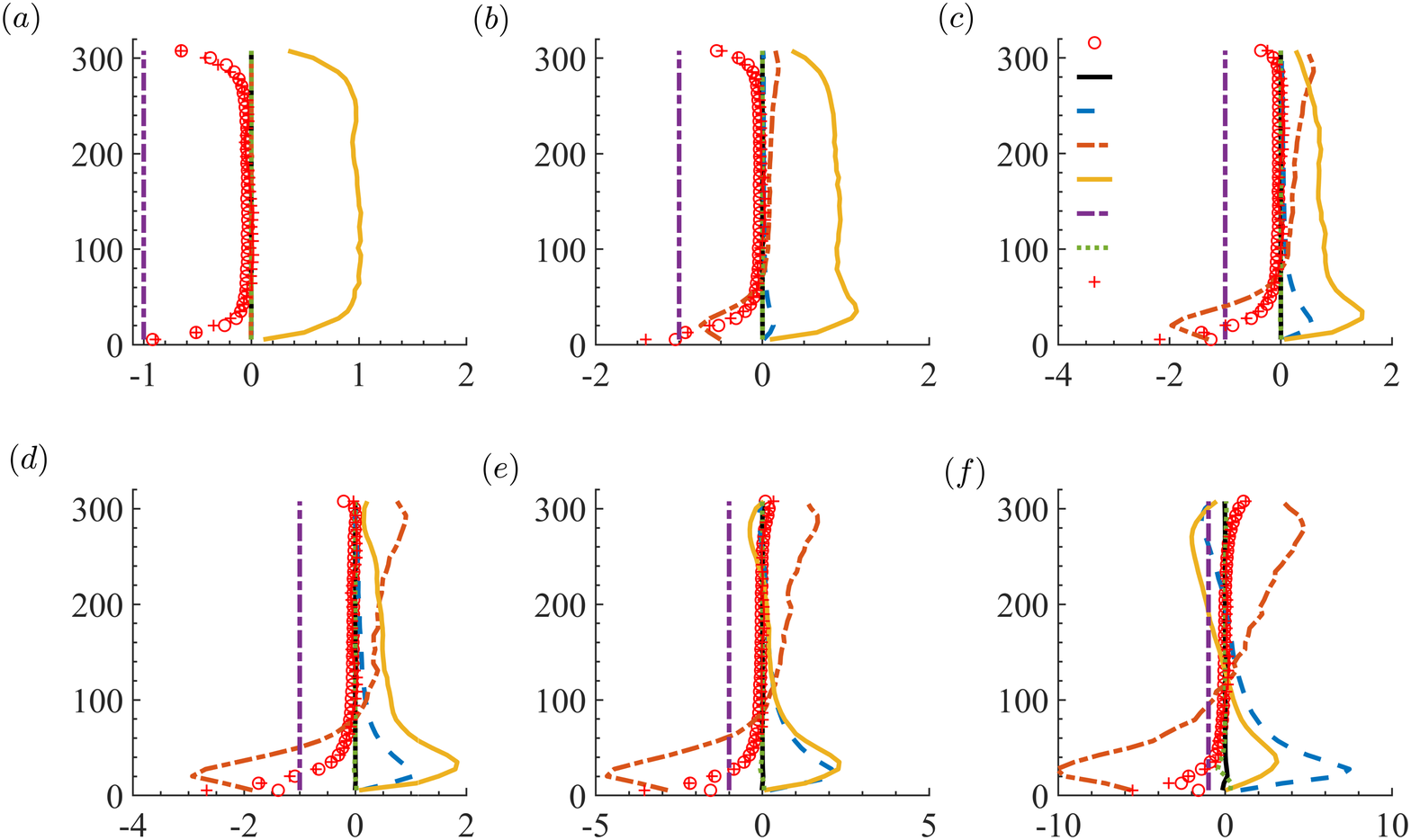}
\put(155,-5){Velocity contribution$/\tau_p g$}
\put(296,210){\scriptsize{{$\langle w^p(t)\rangle_z$}}}
\put(300,201){\scriptsize{{R1}}}
\put(300,192){\scriptsize{{R2}}}
\put(300,183){\scriptsize{{R3}}}
\put(300,174){\scriptsize{{R4}}}
\put(300,165){\scriptsize{{R5}}}
\put(300,156){\scriptsize{{R6}}}
\put(298,148){\scriptsize{{$\sum_i\text{R}_i$}}}
\put(10,165){\rotatebox{90}{$z^+$}}
\put(132,165){\rotatebox{90}{$z^+$}}
\put(255,165){\rotatebox{90}{$z^+$}}
\put(10,45){\rotatebox{90}{$z^+$}}
\put(130,45){\rotatebox{90}{$z^+$}}
\put(250,45){\rotatebox{90}{$z^+$}}
\end{overpic}
\caption{Results for the averaged vertical particle velocity $\langle w^p(t)\rangle_z$, compared with the different contributions to this velocity according to \eqref{rhoeq}, for the zero-flux configuration. Each subplot (a), (b), etc corresponds to Case 1, 2, etc, respectively, and $R_i$ denotes the $i^{th}$ term on the rhs of \eqref{rhoeq}.}
\label{fig:flux}
\end{figure}

In order to understand the physical mechanisms governing the behavior of $\langle w^p(t)\rangle_z$ and $\varrho$, we compute the various terms that contribute to $\langle w^p(t)\rangle_z$ according to \eqref{rhoeq}. Figure~\ref{fig:flux} shows the results for the zero-flux case. Throughout most of the domain where $\langle w^p(t)\rangle_z=0$, we find that for Cases 0 and 1, $\langle w^p(t)\rangle_z\approx \langle u^p(t)\rangle_z-\tau_p g$, which is the behavior expected for a quasi-homogeneous flow according to \eqref{rhoeqQH}. However, the results in figure~\ref{fig:Slog} show that over this same region, the vertical fluid Reynolds stress $\langle u u\rangle$ varies appreciably. This may be understood by noting that in the limit $St^+\to0$, with $Sv^+$ finite, \eqref{rhoeqQH} also reduces to the result  $\langle w^p(t)\rangle_z= \langle u^p(t)\rangle_z-\tau_p g$. For larger $St^+$, the inhomogeneity does play a role, and for Cases 2-5 $\langle w^p(t)\rangle_z\approx \langle u^p(t)\rangle_z-\tau_p g$ does not hold because a significant contribution arises from the turbophoretic velocity $-\tau_p\nabla_z S$ in \eqref{rhoeq} (term R3). This turbophoretic velocity switches from being positive in the upper portion of the domain to negative in the lower portion due to the sign of $\nabla_z S$, whose sign changes because of the change of sign in the gradient of the fluid Reynolds stress (see figure~\ref{fig:Slog}). This means that in the upper portion of the domain, both the turbophoretic velocity and the velocity arising from preferential sampling of the fluid, i.e $\langle u^p(t)\rangle_z$, act against the Stokes settling velocity $-\tau_p g$ in order to preserve $\Phi=0$. Close to the wall where $-\tau_p\nabla_z S$ changes sign and causes particles to drift towards the wall, $\langle u^p(t)\rangle_z$ increases in magnitude, and a diffusion contribution from $(\tau_p/\varrho) S\nabla_z \varrho$ (term R2) is also activated that preserves $\Phi=0$. This diffusion contribution becomes increasingly important as $St$ is increased, as expected based on the discussion in \S\ref{NWR}, and consistent with the results in \cite{johnson20}. For all cases, we find that the contribution from the acceleration $-(\tau_p/2)\nabla_z \langle w^p(t)\rangle^2_{{z}}$ (term R1) and molecular diffusion $(\tau_p\kappa/\varrho)\nabla_z^2\varrho\langle{w}^p(t)\rangle_{z}$ (term R6) terms are negligible, even close to the wall.

\begin{figure}
\centering
\begin{overpic}
[width=0.99\textwidth]{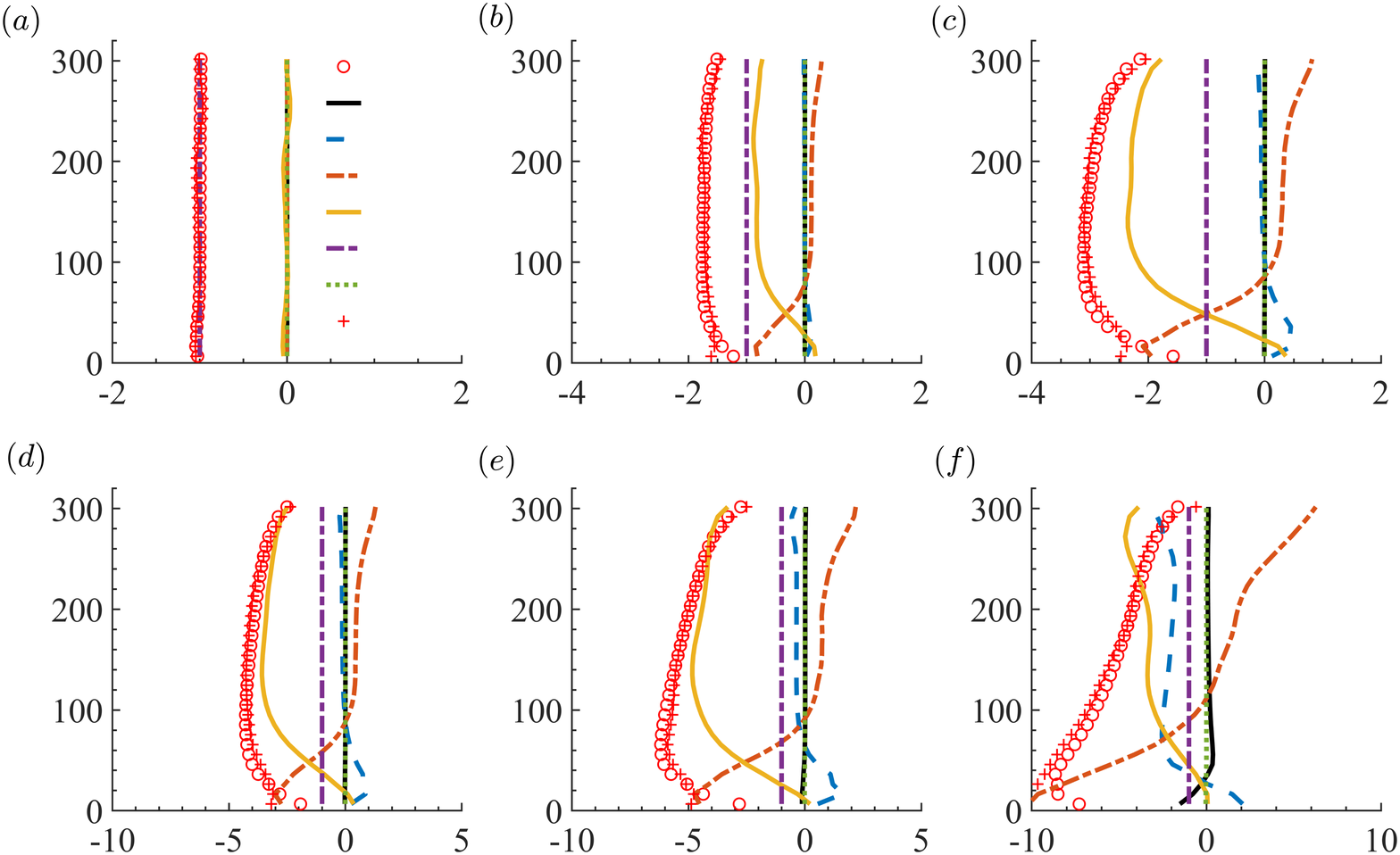}
\put(155,-5){Velocity contribution$/\tau_p g$}
\put(100,216){\scriptsize{{$\langle w^p(t)\rangle_z$}}}
\put(102,206){\scriptsize{{R1}}}
\put(102,196){\scriptsize{{R2}}}
\put(102,186){\scriptsize{{R3}}}
\put(102,176){\scriptsize{{R4}}}
\put(102,166){\scriptsize{{R5}}}
\put(102,156){\scriptsize{{R6}}}
\put(102,146){\scriptsize{{$\sum_i\text{R}_i$}}}
\put(2,175){\rotatebox{90}{$z^+$}}
\put(126,175){\rotatebox{90}{$z^+$}}
\put(249,175){\rotatebox{90}{$z^+$}}
\put(5,55){\rotatebox{90}{$z^+$}}
\put(127,55){\rotatebox{90}{$z^+$}}
\put(247,56){\rotatebox{90}{$z^+$}}
\end{overpic}
\caption{Results for the averaged vertical particle velocity $\langle w^p(t)\rangle_z$, compared with the different contributions to this velocity according to \eqref{rhoeq}, for the constant flux configuration. Each subplot (a), (b), etc corresponds to Case 1, 2, etc, respectively, and $R_i$ denotes the $i^{th}$ term on the rhs of \eqref{rhoeq}.}
\label{fig:flux2}
\end{figure}

In figure~\ref{fig:flux2} we similarly compute for the constant-flux configuration the various terms that contribute to $\langle w^p(t)\rangle_z$ according to \eqref{rhoeq}. Unlike the zero-flux case, for the constant-flux case the particles have a finite average vertical settling velocity. For $St^+\to 0$, $\langle w^p(t)\rangle_z/\tau_p g\to 1$, while for finite $St^+$, $\langle w^p(t)\rangle_z/\tau_p g$ attains values of up to 10, indicating remarkably strong enhancements of the average particle settling speeds due to the combined effects of turbulence and particle inertia (recall that $\langle w^p(t)\rangle_z/\tau_p g=1\, \forall St^+$ in the absence of turbulence). For $z^+>O(100)$, the dominant cause of the enhanced settling velocity comes from $\langle u^p(t)\rangle_z$ (term R4). As discussed in \S\ref{QHR}, when $\Phi<0$ and the flow is homogeneous, $\langle u^p(t)\rangle_z$ is finite due to the preferential sweeping mechanism \citep{maxey1987gravitational,tom19}. However, as explained in \S\ref{AFS}, for wall-bounded turbulence, there is an additional contribution to $\langle u^p(t)\rangle_z$ arising from the combined effects of particle inertia and turbulence inhomogeneity. This additional contribution may explain why we observe larger values for $\langle w^p(t)\rangle_z/\tau_p g$ at $z^+>O(100)$ than have previously been observed for homogeneous turbulence in either DNS where $\langle w^p(t)\rangle_z/\tau_p g\lesssim 2$ \citep{bec14b,ireland16b} or experiments where $\langle w^p(t)\rangle_z/\tau_p g \lesssim 2.7$ \citep{petersen19}. 

As the wall is approached, $\langle u^p(t)\rangle_z$ (term R4) begins to reduce in magnitude (since $u^p(t)\to 0$ for $z^p(t)\to 0$), while the turbophoretic velocity $-\tau_p\nabla_z S$ (term R3) suddenly grows in magnitude, and dominates $\langle w^p(t)\rangle_z$ close to the wall. It is the contribution from $-\tau_p\nabla_z S$ that enables $\langle w^p(t)\rangle_z$ to remain finite as the wall is approached. Physically, the inertial particle remembers its interaction with the turbulence along its path-history in regions where the TKE is finite, and this enables $w^p(t)$ to be finite even if $u^p(t)=0$, such as at the wall. It is this path-history effect that is described by $-\tau_p\nabla_z S$, as explained in \S\ref{NWR}. These results therefore show that as the wall is approached, the importance of the preferential sweeping mechanism in determining the particle settling velocity gives way to the turbophoretic drift mechanism.

Comparing figure~\ref{fig:flux} with figure~\ref{fig:flux2}, we see that in both cases, near the wall the dominant negative contribution to $\langle w^p(t)\rangle_z$ comes from the turbophoretic drift (unless $St^+$ is very small), and that $-\tau_p\nabla_z S$ attains a peak magnitude near the wall that is similar for both cases. The main difference between the two cases concerns the behavior of the positive contributions to $\langle w^p(t)\rangle_z$. In particular, for the constant-flux configuration, the absorbing wall boundary condition means that once the particles have reached the wall, they do not have enough time close to the wall in order to experience sufficiently large positive values of $u^p(t)$ that can transport them away from the wall. This differs from the zero-flux case for which $\langle u^p(t)\rangle_z>0$ near the wall enabling the particles to be suspended back into the flow from the near wall region, producing the zero-flux state. The diffusion term $-\tau_p S\nabla_z\varrho$ (term R2) is also much smaller in the near wall region for the constant flux case than it is for the zero-flux case.

Similar to the zero-flux configuration, for the constant-flux configuration we find that the contribution from the acceleration $-(\tau_p/2)\nabla_z \langle w^p(t)\rangle^2_{{z}}$ is negligible, even close to the wall. Therefore, for both configurations, the first and sixth terms on the rhs of \eqref{rhoeq} (R1 and R6) may be neglected (R6 is identically zero for our constant flux case), justifying this assumption in the analysis of \S\ref{Asymp}.

\FloatBarrier

\subsection{Extending the Rouse model}

As stated in the introduction, one of the motivations for our study is to consider how the Rouse model for particle concentration (that was derived for $St^+\to 0$, with finite $St^+/Fr^+$) must be extended in order to apply for $St^+\geq O(1)$. Based on the results in this section, terms R1 and R6 in \eqref{rhoeq} may be neglected. Furthermore, if we use \eqref{uclose} to model $\langle{u}^p(t)\rangle_{{z}}$, then for the zero-flux case we obtain the approximate version of \eqref{rhoeq}
\begin{align}
0\approx -(\tau_p S+\mathcal{D}^{[1]}+\kappa) \nabla_z \varrho - \varrho\Big(\tau_p\nabla_z S -\zeta+\tau_p g\Big).\label{rhoeq3}
\end{align}
Comparing this with the Rouse model in \eqref{Rouse_rho} reveals a number of differences, and points to the way in which the Rouse model is to be extended to apply for $St^+\geq O(1)$. First, Rouse's eddy-diffusion model $\mathcal{K}$ that only applies in the log-law region can be replaced with the more general diffusion coefficient $\mathcal{D}^{[1]}$ that is valid for arbitrary $z^+$, and for which a simple closed expression is given in \eqref{Dform}. Second, an additional contribution to the diffusion coefficient must be accounted for, namely $\tau_p S$, which captures the diffusion contribution arising because of the imperfect coupling between the fluid and inertial particle velocities. Third, the turbophoretic velocity $-\tau_p\nabla_z S$ must be accounted for. Fourth, the drift contribution $\zeta$ must be accounted for that captures the effects of the preferential sweeping mechanism of \cite{maxey1987gravitational}, as well as preferential sampling of the flow due to turbulence inhomogeneity. The study of \cite{RichterBLM2018} captures these additional effects for $St^+\ll1$, but does not apply for $St^+\geq O(1)$.

The terms involving $S$ are unclosed, and $S$ must be predicted, yet its transport equation is unclosed. Models such as \cite{zaichik99} attempt to close these transport equations using a quasi-normal approximation. While this may lead to reasonable results far enough away from the wall, near the wall such a closure is known to yield behavior that is inconsistent with the behavior predicted using asymptotic analysis (see \cite{sikovsky14}). Developing closures that are consistent with this asymptotic behavior is crucial since it is in the near wall region where most of the complexity in the particle motion occurs, e.g. where the strong particle accumulation occurs. It is also necessary to test the accuracy of the closure in \eqref{uclose}, and to develop a closed form expression for the drift velocity $\zeta$. Consideration of these issues will be the subject of our future work.

\section{Conclusions}

We have used a combination of theoretical analysis and DNS data to explore the mechanisms and behavior of the settling velocities and spatial distributions of inertial particles in a wall-bounded turbulent flow. Two different flow configurations were considered, one where the particle mass flux is zero, and the other where it is  constant and negative. 

The theory is based on the exact transport equations for the particle statistics that are derived from a phase-space, master PDF equation. This allowed us to identify and consider the specific contribution to the particle settling velocities and spatial distribution coming from distinct physical mechanisms in the system. We then examined the asymptotic behavior of the particle motion in the near-wall region where the particle accumulation is strongest. The results revealed the fundamental differences to the near wall behavior of the particle statistics that is produced by gravitational settling, compared to the no-settling case that was previously explored in \citet{sikovsky14} and \citet{johnson20}. We also identified a regime where the particle concentration grows as a power-law in the near-wall region, for which the power-law exponent increases with increasing $St^+$ for $St^+\geq O(1)$. This regime was not identified in the previous analysis of \cite{sikovsky14} because in that study the limit $z^+\to 0$ was taken rather than simply considering the regime $z^+\ll 1$, or in \cite{johnson20} which only considered $St^+\ll 1$.

For the zero-flux case, the DNS results revealed that the vertical particle motion is similar to the behavior without gravitational settling. The particle concentration grows as a power-law as the wall is approached, which is described by the new power-law regime just discussed, rather than that described in \cite{sikovsky14,johnson20}. In our simulations, a diffusion term was added to the particle motion in order to enable them to escape the wall where they are introduced, and for $z^+\leq O(10)$ this dominates the behavior and causes the concentration to become constant, and the average vertical particle velocity to become finite. For the constant flux case, the combined effects of turbulence and particle inertia lead to average vertical particle velocities that can significantly exceed the Stokes settling velocity. In particular, as the particles approach the wall, their average vertical velocity can significantly increase, depending on $St^+$, reaching values up to ten times the Stokes settling velocity. Below a certain $z^+$, however, the average vertical particle velocities reduce due to the reduction of the fluid velocities as the wall is approached. 

Concerning the mechanisms governing the average vertical particle velocity, in the zero-flux case, at heights $z^+\geq O(100)$, the average velocity is zero, and for the lower $St^+$ cases this is due to a downward contribution from the Stokes settling velocity that is precisely balanced by an upward velocity arising from the particles preferentially sampling regions of the flow where the fluid velocity is positive. For larger $St^+$ there is also an upward turbophoretic velocity (since in this region the fluid Reynolds stresses decay with increasing $z^+$) that acts together with the preferential sampling effect to counter balance the Stokes settling velocity. As the particles approach the wall, they experience a strong turbophoretic velocity contribution that drives them towards the wall, that is counteracted by an upward velocity contribution arising from the preferential sampling of regions where the fluid velocity is positive, and additional contributions arising from diffusive mechanisms that are driven by gradients in the concentration field. 

For the constant flux case, for $z^+\geq O(100)$ the average vertical particle velocities can significantly exceed the Stokes settling velocity due to the particles preferentially sampling regions where the fluid velocity is negative. This effect is associated with the preferential sweeping mechanism of \cite{maxey1987gravitational}. As $St^+$ increases, there is also an upward contribution from the turbophoretic velocity, but this is overwhelmed by the contribution from preferential sweeping. As the particles approach the wall, the contribution to the average vertical particle velocity coming from the preferential sweeping mechanism becomes small, and a downward contribution from the turbophoretic velocity dominates the behavior.

For future work, it is important to consider how the behavior observed here changes when $Sv^+$ is varied, since this quantity was held fixed in our simulations in order to isolate the effect of $St^+$. In the environment, $St^+$ and $Sv^+$ will vary simultaneously, and as such, different mechanisms may compete and play dominant roles compared with the case we have explored. It will also be interesting to perform DNS using more particles, and/or longer simulation times in order to generate robust statistics very close to the wall so that the asymptotic predictions we have derived may be explored more thoroughly. Finally, one of the motivations for this study was to better understand the role of particle inertia in order to understand how the Rouse model for the particle concentration, which was derived for $St^+\to 0$ (with finite $St^+/Fr^+$), can be modified for $St^+\geq O(1)$. Our study, and results from the future research just discussed can provide crucial insights guiding the particular terms and mechanisms that must be incorporated into such an extended Rouse model. For example, our present results show that in order for the Rouse model to describe the regime $St^+\geq O(1)$, it must be extended to include the turbophoretic drift velocity, a diffusion mechanism associated with the inertial particle velocities being partially de-coupled from the local fluid velocity, as well as the term describing the preferential sampling of the fluid velocity field, which captures the preferential sweeping mechanism.

\section*{Acknowledgements}

The authors acknowledge grant G00003613-ArmyW911NF-17-0366 from the US Army Research Office. Computational resources were provided by the High Performance Computing Modernization Program (HPCMP), and by the Center for Research Computing (CRC) at the University of Notre Dame.

\section*{Declaration of Interests}

The authors report no conflict of interest.

\bibliographystyle{jfm}
\bibliography{bib/reference_combine}

\end{document}